\documentclass[journal]{IEEEtran}
\usepackage{graphicx}

\ifCLASSINFOpdf
\else
 \fi
\begin{document}\title{The Role of the Critical Ionization Velocity Effect in Interstellar Space and the Derived Abundance of Helium}

\author{Gerrit L. Verschuur,~\IEEEmembership{}
        Joan T. Schmelz,~\IEEEmembership{}
        and~Mahboubeh Asgari-Targhi,~\IEEEmembership{}
% <-this % stops a space
\IEEEcompsocitemizethanks{\IEEEcompsocthanksitem G. Verschuur is retired, unaffiliated and living in Siunnyvalre, CA.\protect\\
E-mail: verschuur@aol.com\protect\\
J. Schmelz is with USRA, Mountain View, CA.\protect\\
 E-mail:jschmelz@usra.edu \protect\\
M. Asgari-Targhi is at the Center for Astrophysics, Cambridge, MA.\protect\\
E-mail: Mahboubeh Asgari-Targhi@cfa.edu
}% <-this % stops a space
\thanks{Manuscript received ; revised .}}

%markboth{Journal of \LaTeX\ Class Files,~Vol.~14, No.~8, August~2015}%

\maketitle

\ifCLASSOPTIONcaptionsoff
  \newpage
\fi

\begin{abstract}
Gaussian analysis of new, high-angular-resolution interstellar 21-cm neutral hydrogen emission profile structure more clearly reveals the presence of the previously reported signature of the critical ionization velocity ({\it CIV}) of Helium (34 km s$^{-1}$).  The present analysis includes 1496 component line widths for 178 neutral hydrogen profiles in two areas of sky at galactic latitudes around $-$50$^\circ$, well away from the galactic plane. The new data considered here allow the interstellar abundance of Helium to be calculated, and the derived value of 0.095 $\pm$ 0.020 compares extremely well with the value of 0.085 for the cosmic abundance based on solar data.  Although the precise mechanisms that give rise to the {\it CIV} effect in interstellar space are not yet understood, our results may provide additional motivation for further theoretical study of how the mechanism operates.
\end{abstract}

% Note that keywords are not normally used for peer review papers.
\begin{IEEEkeywords}
Critical Ionization Velocity, Milky Way, Interstellar Matter, Cosmic Abundances
\end{IEEEkeywords}

\IEEEpeerreviewmaketitle

\section{Introduction}
Verschuur \& Peratt (1999) [1] and Peratt \& Verschuur (2000) [2] noted that interstellar neutral hydrogen emission line profiles revealed the signatures of the Critical Ionization Velocity ({\it CIV}) of the most abundant interstellar species; viz, Helium (He) at 34.3 km s$^{-1}$, part of Band I (Peratt 2015) [3], and the four elements Carbon (C), Nitrogen (N), Oxygen (O), and Neon (Ne) that cluster around 13 $-$ 14 km s$^{-1}$, which make up Band II.  Verschuur \& Schmelz (2010) [4] subsequently used a great deal more data and argued strongly that the observed 34 km s$^{-1}$ wide HI emission line component found in their work as well as in the line width values published by nine other authors is related to the {\it CIV} of helium.  It is difficult to observe helium directly in interstellar space, which means that the presence of 34 km s$^{-1}$ wide neutral hydrogen emission line component opens a new window into deriving the composition of interstellar matter.  

\section {Analysis}

The data used here are from the 2nd data release of the Galactic Arecibo L-Band Feed Array ({\it GALFA}) neutral hydrogen (HI) survey [5].\footnote[1]{The Inner Galaxy ALFA (I-GALFA (http://www.naic.edu/~igalfa) survey data are part of the Galactic ALFA ({\it GALFA}) H I project observed with the Arecibo L-band Feed Array (ALFA) (beam width 4 minutes of arc) on the 305-meter William E. Gordon Telescope. The Arecibo Observatory is a U.S. National Science Foundation facility that, at the time, was operated under a Cooperative Agreement with SRI International, Universities Space Research Association (USRA), and Universidad Metropolitana.}\footnote[2] {The data can be downloaded for small areas of sky at https://purcell.ssl.berkeley.edu/cubes.php.}  The channel bandwidth of the data used here is 0.7 km s$^{-1}$. The brightness temperature noise is 140 mK r.m.s. per 1 km s$^{-1}$. The survey included all the neutral hydrogen in and around the Milky Way galaxy in the 0$^\circ$ $-$ 40 $^\circ $declination range.  

The angular resolution of these data is a factor of 10 greater than that of earlier 21-cm surveys used to investigate the relationship between HI emission line components and {\it CIVs}. 

Gaussian decomposition provides information on the properties of each component including the peak brightness temperature, T$_{B}$, line width, W, center velocity, V$_{c}$, and the area of the Gaussian, which gives its total hydrogen column density along the line-of-sight, N$_{H}$.  These parameters provide information on the physical conditions of the astrophysical environments producing the HI profiles in any given direction. For example, the line widths themselves are generally assumed to reveal the kinetic temperature of the gas, but we will argue here that this may not always be the case.

For the present study, Table I lists details of the directions studied.  Right Ascension (RA) is shown in column 1, the Declination in column 2, and the corresponding galactic coordinates, longitude and latitude, in columns 3 \& 4.  These are the central coordinates of each series of profiles separated by 0.$^\circ$1\  in Declination at the RA shown.  The HI profiles over the full velocity range for these directions were decomposed into Gaussians.  For each direction, the number of profiles listed in column 5 were decomposed into the number of Gaussians listed in column 6.  The average reduced chi-squared values for the best fit solution (discussed below) together with the standard deviation are shown in columns 7 \& 8.  The first set of 6 entries in Table I includes cuts across a long filamentary feature of hydrogen gas in the southern galactic sky moving at $-$60 km s$^{-1}$ with respect to the observer, which has been fully mapped by Verschuur et al. (2018) [6].  A second set involves cuts across an HI spheroidal feature just to the north of this filament.

The analysis of the 21-cm emission profile shapes was carried out using the Solver algorithm (see, e.g., Verschuur 2004 [7]).  Previous detailed analysis of these regions by Verschuur et al. (2018) [6] has shown that a broad underlying component is present in all directions, as revealed in low-level wings in the overall profile. 
The Solver method require a first guess for the Gaussian parameters. Based on the Verschuur et al. (2018) [6] analysis, we begin with one Gaussian component that fits the wings of the profile. This component is removed, and additional components are added if the residuals from the previous iteration still show significant structure. Solver is then run on this initial guess and allowed to fit as many as 10 components, which is seldom required. The best-fit solution is defined as one in which the square root of the sum of squares of the residuals over the velocity range of the emission profile is a minimum. 

Solver will fit additional components as required or fewer, if that produces a best fit.  At this stage our method takes a further step compared to previous studies.  The reduced chi-squared ($\tilde{\chi}^{2}$) value over the velocity range of the profile of the apparent best fit is calculated.  If $\tilde{\chi}^{2}\ <\ 0.8$, the solution may contain too many components.  Examination of the profile and the residual of the fit usually shows where a superfluous component is present. The superfluous component is removed, and the algorithm is again run again, often resulting in a solution with $\tilde{\chi}^{2}\ \approx\ 1.0$. If, on the other hand, $\tilde{\chi}^{2}\ >\ 1.2$, another component may be needed to produce a better fit. At this stage the residual usually indicates where such a component is likely to be located.  The algorithm is then again run and the result for reduced chi-squared is usually close to unity.

The solution found for the first profile in any given data set is used as the starting point for the next adjacent profile, and so on. This technique, which was originally developed for simple profiles that could be fit with a few components, can readily be extended for more complex profiles such as those seen in Fig. 1. The panels highlight two spectral components that are present in all directions $-$ a broad underlying component of order 34 km s$^{-1}$ wide (shown in red) that fits the profile wings, and a second component of order 14 km s$^{-1}$ wide (shown in green). Other narrower components are superimposed on these two. The underlying Gaussian component for the negative velocity gas invariably has line widths of order 21 km s$^{-1}$ which shows no hint of components of order 34 or 14 km s$^{-1}$ wide in this gas.

\begin{figure*}
		\centering
	\includegraphics[width=0.8\textwidth]{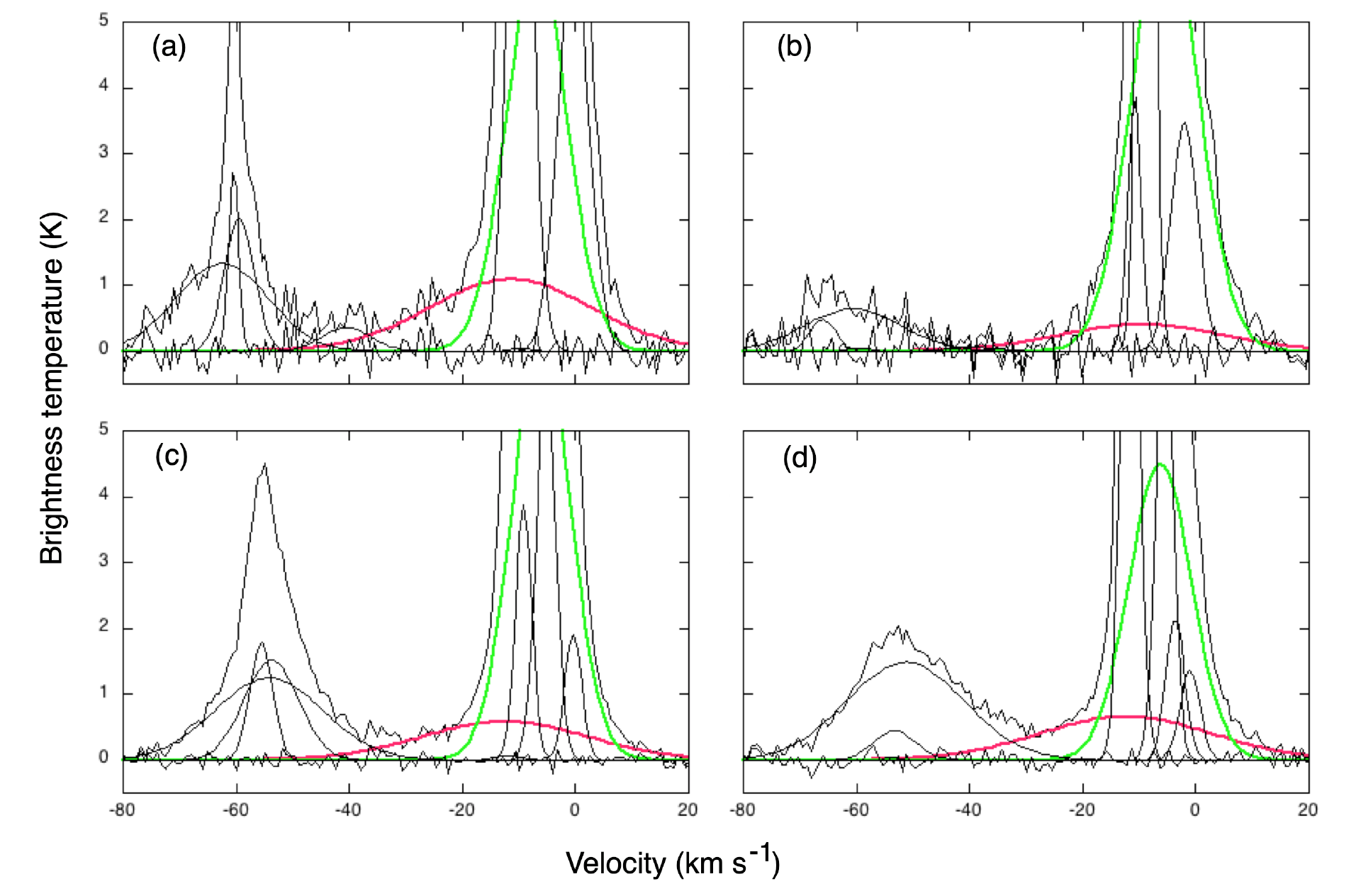}
	\caption{Examples of the Gaussian fits for four profiles showing the underlying broad component of order 34 km s$^{-1}$ wide in red and a second pervasive component of order 14 km s$^{-1}$ wide in green. The narrow components are all shown as black solid lines.  The residuals found after Gauss fitting are shown.  In all cases the Solver algorithm was set to fit a maximum of 10 components.  The profiles for the southern filamentary feature are for (a) at Right Ascension 360.$^\circ$0, Declination 8.$^\circ$7 and for (b) at 358.$^\circ$0, 9.$^\circ$9.  The peaks around $-$60 km s$^{-1}$ arise in the southern filamentary feature.   Two profiles for the southern spheroidal feature are shown in (c) at Right Ascension 1.$^\circ$6, Declination 11.$^\circ$7 and in (d) at 2.$^\circ$1 and 11.$^\circ$2.  The peak around $-$50 km s$^{-1}$ pertains to the southern spheroidal feature, which is again separate from the low velocity peak.  The brightness temperature scale on the vertical axis is chosen to detail the Gauss fitting rather than to display the full amplitude of the entire profile.}

\end{figure*}

The present study included Gaussian analysis of 178 individual HI profiles that netted 1496 components. For the final analysis, components with peak brightness temperatures $<0.1$ K and/or line widths $<1.5$ km s$^{-1}$ are removed from further consideration since they are at the level of noise on the data.  

The histogram in Fig. 2 combines the line width results for all the components derived from the Gaussian fit to the HI profile data for the sample given in Table I. The peak around 34 km s$^{-1}$ is similar to the results from the literature as summarized by Verschuur \& Schmelz (2010) [4].

\begin{figure}
\centering
\includegraphics[width=0.4\textwidth]{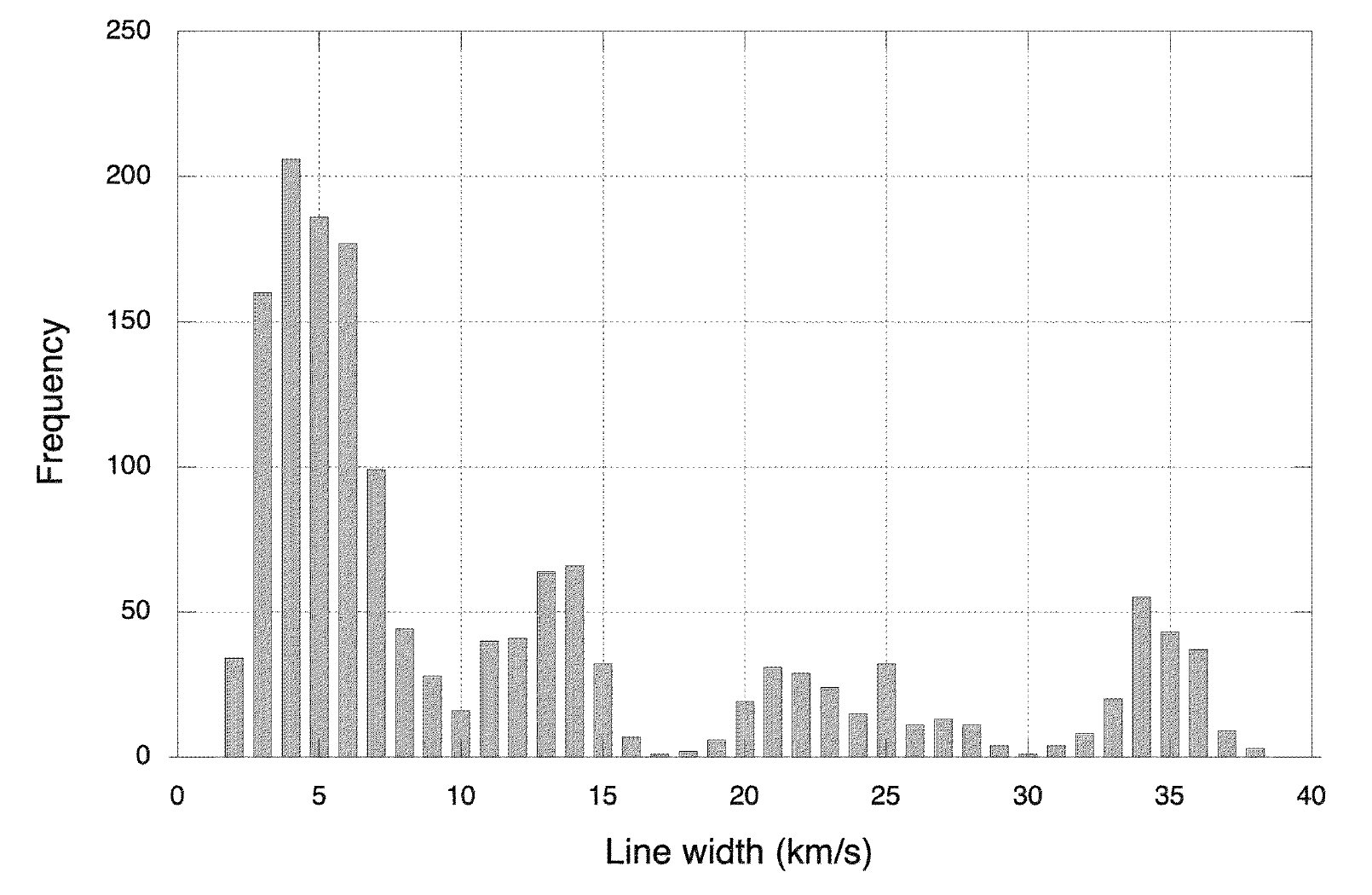}
\caption{The 21-cm interstellar hydrogen line component line width distribution for 1496 Gaussians derived from 178 profiles in two areas of sky using {\it GALFA} HI survey data from the Arecibo 305-m diameter radio telescope. The clustering is related to the {\it CIVs} of the most abundant interstellar species: Helium around 34 km s$^{-1}$ and neutral C, N, O \& Ne around 14 km s$^{-1}$ . The large peak around 4 km s$^{-1}$ includes contributions from cold (100 to 300 K) HI and possibly the {\it CIV} signatures of heavier elements such as Na, Ca and Fe in the range 5 to 8 km s$^{-1}$.  The broad distribution from 20 $-$ 30 km s$^{-1}$ is produced solely by the HI peaks at negative velocities seen in Fig. 1. }
\end{figure}

The traditional interpretation of a Gaussian line width, $W$, is that it indicates the kinetic temperature, T$_{K}$ [8], [9] of the gas involved, which makes the component at 34 km s$^{-1}$ particularly intriguing. Using T$_k$ $=$ 21.86 W$^{2}$ K [10] implies a temperature of 24,000 K, high enough to ionize the hydrogen gas so the 21-cm HI emission spectra should not be visible. This concern originally led to the speculation [11] that such a component might result from side-lobe contamination. However, using side-lobe-corrected data [7], [1] from the Leiden/Dwingeloo HI survey [12], Gaussian analysis still revealed a pervasive, underlying, 34 km s$^{-1}$ wide component in all directions. Turbulence could explain a broad component, but there is no reason to expect that it should always have the same numerical value. 

Our new data now confirm the presence of a pervasive broad Gaussian component with a line width of order 34 km s$^{-1}$, which may be explained by the {\it CIV} of helium.

There is (to our knowledge) only one result in the refereed literature that disagrees in part with the histogram in Fig. 2. Although the distribution in Fig. 3b of Nidever et al. (2008) [9] contains a secondary peak at 34 km s$^{-1}$, it is not sharply peaked like our Fig. 2. Rather, it has a long tail extending to higher velocities revealing a continuum of very broad line components, which might not be consistent with the CIV model.

Verschuur\& Schmelz (2020) [13] were able to examine the Nidever et al. (2008) [9] Gaussian fits in detail and compare the results of their automated IDL technique with our profile-by-profile approach. Both methods use data from the Leiden/Argentine/Bonn all-sky survey. The comparisons led to the identification of four problems with the Nidever et al. (2008) [9] analysis: (1) different methods of calculating the $\tilde{\chi}^{2}$ measuring the goodness of fit; (2) an ultra-broad component found bridging the gap between low- and intermediate-velocity gas; (3) multiple, fundamentally different solutions for the profiles at both the north and south galactic poles; and (4) the lack of an imposed spatial coherence allowing different components to appear and disappear in profiles separated by a fraction of a beam width. Verschuur\& Schmelz (2020) [13] were able to show that the long tail in the distribution in Fig. 3b of Nidever et al. (2008) [9] disappeared once these corrections were made to the Gaussian fitting procedure and that the modified results appear to be consistent with Fig.2 and the CIV model.

\section{On the nature of the CIVs}

The 34 km s$^{-1}$ component may be associated with the plasma phenomenon known as the {\it CIV} [1], which is defined as that velocity at which a neutral particle traveling through a plasma and normal to the magnetic field becomes ionized. That occurs when its kinetic energy is equal to its ionization potential. The phenomenon was first proposed by Alfv\'{e}n [14], [15] and has since been extensively verified in terrestrial laboratory and near-Earth space-borne tests, as reported in a number of reviews [16], [17], [18], [19].

The critical velocity, V$_{cr}$, is defined by:
\begin{equation}
V_{cr} = (2eV/M)^{0.5} 
\end{equation}
where M is the mass of the atom (or ion) and eV is its ionization potential. 

As pointed out in these reviews, the precise mechanism that drives the {\it CIV} effect remains elusive. Peratt \& Verschuur (2000) [2] invoked the plasma two-stream instability. A stream of energetic electrons passing through cold plasma excites ion waves, which will grow rapidly in magnitude at the expense of the kinetic energy of the electrons. A high energy tail of the electron distribution ionizes the neutrals.  Others [18], [19] emphasized the importance of lower hybrid instabilities in facilitating the ionization.  Although researchers have not settled on a theoretical explanation that satisfies everyone, the observational evidence for the {\it CIV} effect occurring in nature seems to be beyond question. 

The neutral gas masses and plasma observed within the beam of a radio telescope encompass large volumes of space and the subsequent {\it CIV}-induced motions will modulate the 21-cm hydrogen spectral lines whose widths are then determined by the {\it CIV} effect acting on a range of atomic species. For example, if the motions of the ionizing electrons (coupled to the magnetic field and the neutrals) extend between + and - 34 km s$^{-1}$ along the line-of-sight, the Gaussian line widths we observe would appear to have a width at half maximum of order 34 km s$^{-1}$.

The outstanding thing about the 34 km s$^{-1}$ component is that it cannot be explained with the traditional kinetic temperature model. This is because the emission line width would imply a temperature high enough to ionize the very gas that is producing the observed neutral hydrogen profile. Although the narrower peaks in Fig. 2 can be interpreted as temperature, they can also result from the {\it CIV} effect for other astrophysically prominent atoms. For example, the peak in Fig. 2 around 14 km s$^{-1}$ may correspond to the {\it CIV} band of C, N, O, and Ne, the most abundant elements in the Galaxy after H and He. The significance of those results will be considered in a later paper.

The {\it CIV} effect in interstellar space will not operate if it exceeds the Alfv\'{e}n velocity in the medium.  For the two areas discussed here, which are at large galactic latitudes ($-$50$^\circ$), it is likely that the HI gas is located in the thick disk where interstellar electron densities are low, of order 0.01 to 0.03 cm$^{-3}$, certainly less than the value of 0.1 cm$^{-3}$ found in the so-called Local Interstellar Cloud (Redfield \& Falcon, 2008 [20]). More specifically, based on pulsar dispersion measure data, Cordes \& Lazio (2018) [21] find 0.01 cm$^{-3}$ as seen in their Fig. 3.  Estimates of the ambient magnetic field strength are sparse though Verschuur et al. (2018) [6] estimated that the magnetic fields involved in controlling the stability of the southern filament lie in the range 5 to 18 microGauss.  These values imply Alfv\'{e}n velocities of order 100 km s$^{-1}$ or greater. Therefore, the {\it CIV} for helium, 34 km s$^{-1}$ is permitted. 

Bear in mind that the interstellar medium is permeated by distorted magnetic fields, which in some areas have a net preferred direction.  In general, however, it is reasonable to assume that there will be deviations in direction on all scales smaller than the beam width. That means that as seen from the point of view of neutral gas or plasma streams interacting throughout this medium, the velocity component of the motion normal to localized field segments will have a large range of values. Given that the {\it CIV} effect depends on the velocity of the neutrals with respect to the plasma (or vice versa) normal to the localized magnetic field direction, a wide range of {\it CIV}s will be triggered in the space enclosed within the beam of the radio telescope used to study the HI emission profile structure. A source of relative motion is required to trigger the {\it CIV} effect, and old supernova expanding shells and/or large-scale accretion of matter from the galactic halo onto the disk are obvious candidates. Thus, in any given direction the {\it CIV} signatures related to the entire range of atomic species may be expected.

Table II summarizes the Gaussian component line width data for the directions toward southern filamentary and spheroidal features.  The right ascension label from Table 1 is listed in column 1 followed by the total hydrogen column density in units of 10$^{18}$ cm$^{-2}$ in column 2 for all the profiles in the sample, noted in Table I, column (5)..  Column 3 gives the total column density of the hydrogen in the 34 km s$^{-1}$ wide components seen in Fig. 2, that is, the hydrogen being affected by the He {\it CIV}.  The average line width for these components and their standard deviation are listed in columns 4 \& 5.   The weighted average line width for entire ensemble of 12 entries in Table II is 34.2 $\pm$ 0.3 km s$^{-1}$, which compares to the {\it CIV} for helium of 34.3 km s$^{-1}$.  We regard this as a great more than just coincidence and therefore worthy of further consideration.  Our next step is to use this result to derive the interstellar He abundance, which is obtained from the ratio of the data in columns 3 \& 2 and listed in column 6.

\section{Discussion}

If we now assume that the {\it CIV} effect operates in interstellar space and determines the clustering of HI component line widths that we see in Fig.2, we can  use this result to investigate the abundance of helium.  This can be done by comparing the column density of the helium signature seen in the 34 km s$^{-1}$ wide component with respect to the total column density of the HI observed along the entire line of sight.  We propose that this ratio is the interstellar helium abundance, which is listed in column 6 of Table II for each of the data samples.  The weighted average helium abundance for all 12 entries in the table is 0.095 $\pm$ 0.020.  This compares extremely well with a value of 0.085, the photospheric/cosmic abundances found in the literature: [22], [23], [24].  Although abundances may deviate from photospheric values in certain astrophysical environments such as the solar corona, solar wind, and solar energetic particles [25], and, on occasion, one element may deviate from the norm [26], [27], photospheric abundances are widely accepted to be good measures of their cosmic values.  Thus the helium abundance derived from the {\it CIV} data is within one sigma of the photospheric/cosmic value. It is difficult to observe helium directly in interstellar space, so the 34 km s$^{-1}$ feature in the 21-cm spectrum opens a new window into investigating the composition of interstellar matter.  

Note that at 51.0 km s$^{-1}$, the {\it CIV} effect ionizes hydrogen atoms, but that signature is not revealed in the 21-cm data because the resulting protons and electrons do not contribute to the observed emission profiles. Similarly, the He$^{+}$ produced by the {\it CIV} effect at 34 km s$^{-1}$ discussed here has a {\t CIV} at 51.1 km s$^{-1}$ so that the once-ionized helium loses its remaining electron to create an alpha particle and an additional electron.  These particles are also lost to our observations.  Both a certain amount of hydrogen and ionized helium effectively $``$disappear$"$, i.e., no longer contribute to the 21-cm emission profile, at 51 km s$^{-1}$, and therefore the column densities of the remaining helium and hydrogen allow the He-to-H ratio to be derived as shown in Table II.

\section{Conclusion}
Analysis of new, high-resolution observations of 21-cm interstellar neutral hydrogen emission profiles from the GALFA-HI survey confirm the results of Verschuur \& Schmelz (1989) [11], who showed that a good fit generally required a broad component with a width in order 34 km s$^{-1}$. Since this component could not be interpreted as the kinetic temperature or result from sidelobes, they suggested that the feature might be due to the {\it CIV} effect operating in interstellar space.  By accepting this conclusion, we can derive a value for the interstellar abundance of helium, 0.095 $\pm$ 0.020, which agrees to within half a standard deviation with the cosmic abundance value of 0.085 [22], [23], [24].  

\section*{Acknowledgment}
We wish to thank Dr J. E. G. Peek and his team for all their great work on the GALFA HI Survey and for making their data cubes so readily available.  GLV would particularly like to thank Dr. A. L. Peratt for his encouragement over many years as GLV skirted the mysteries of plasma physics.

\begin{IEEEbiography}  [{\includegraphics[width=1in,height=1.25in,clip,keepaspectratio]{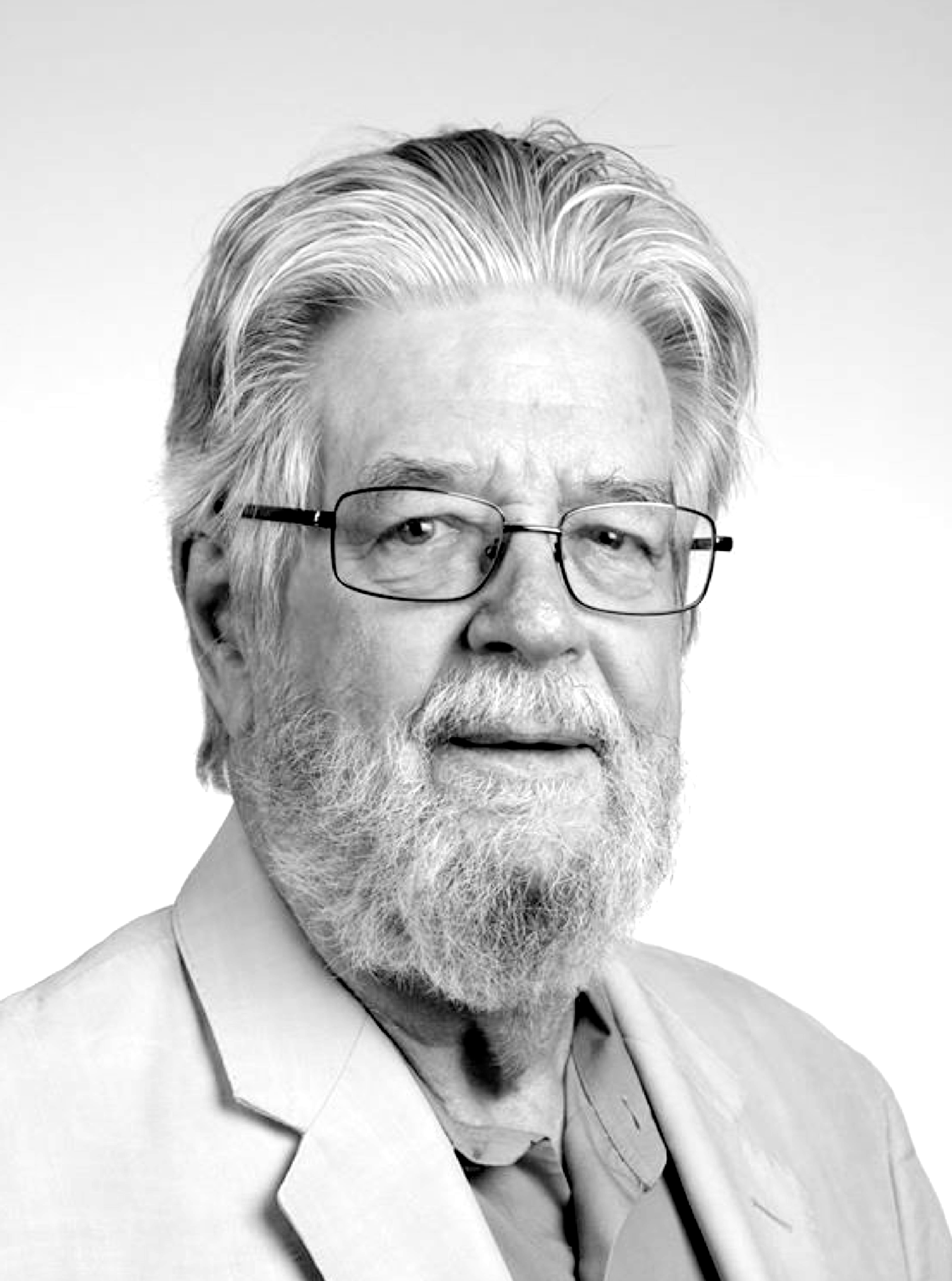}}]{Gerrit Verschuur}
was born in Cape Town and obtained a MSc degree at Rhodes University in South Africa.  His Ph.D.is from the University of Manchester (Jodrell Bank) in England.  After 6 years at  the National Radio Astronomy Observatory in Charlottesville he moved to the University of Colorado in Boulder.  He became a naturalized US citizen in 1975.  He is best known for his work in radio astronomy, has published extensively, and is the author of 12 books.  As an inventor he has a dozen patents to his credit.  In his primary field of study, he pioneered the measurement of the interstellar magnetic field using the 21-cm Zeeman effect technique.  Following subsequent stints at various universities he was for 3 years Emeritus Astronomer at the Arecibo Observatory and is now unaffiliated and living in Silicon Valley.  He continues to carry research into the nature of interstellar neutral hydrogen structure and dynamics, especially with regard to the importance of plasma phenomena in interstellar space.
\end{IEEEbiography}

\begin{IEEEbiography}  [{\includegraphics[width=1in,height=1.25in,clip,keepaspectratio]{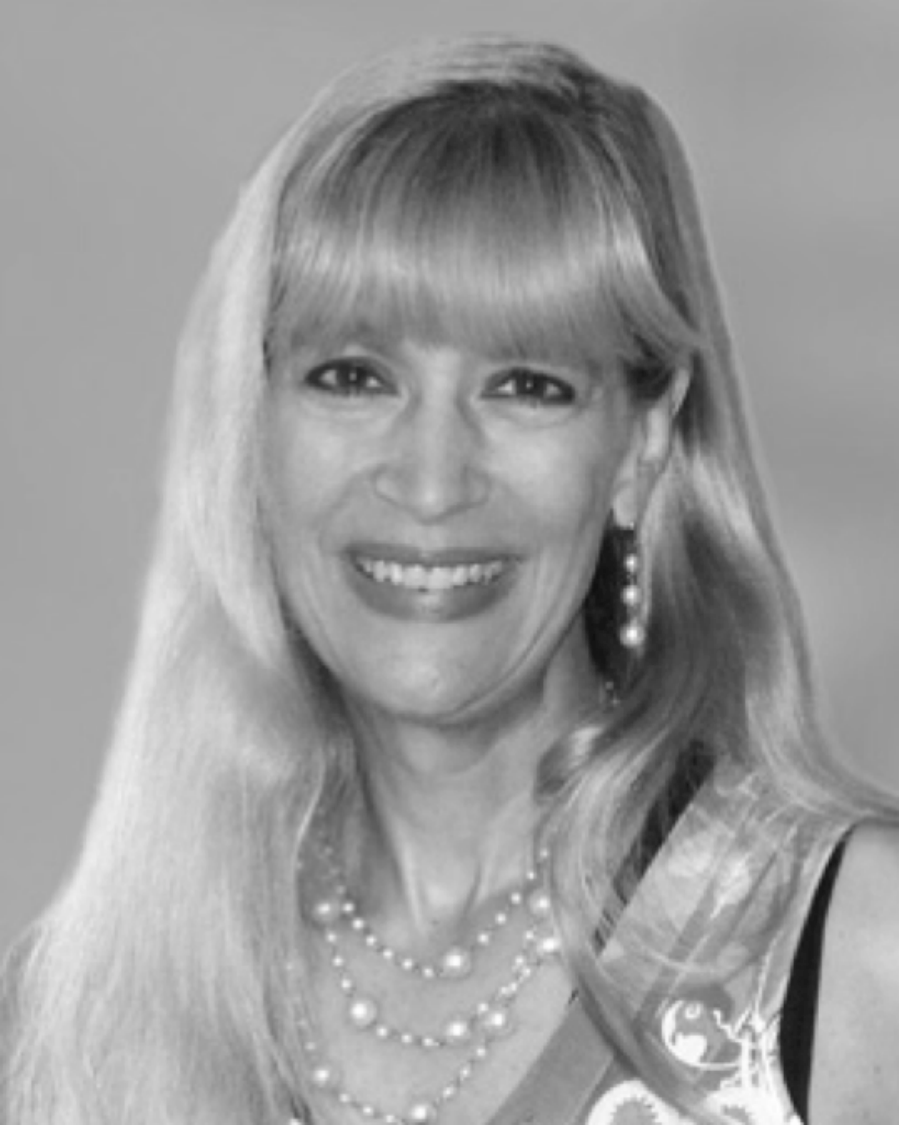}}]{Joan Schmelz}
works for Universities Space Research Association (USRA) and currently serves as the director of the NASA Postdoctoral Program. She is also the Associate Director for Science and Public Outreach for the Stratospheric Observatory for Infrared Astronomy (SOFIA) at NASA Ames Research Center. She is the former deputy director of the Arecibo Observatory in Puerto Rico and a former program officer for the National Science Foundation's Division of Astronomical Sciences. She was a professor at University of Memphis for over 20 years where she worked with dozens of undergraduate students developing observational constraints to test solar coronal heating models. Schmelz has published papers on a variety of astronomical topics including stars, galaxies, interstellar matter, and the Sun using data from ground- and space-based telescopes at (almost) every band of the spectrum.  She is a current Vice President of the American Astronomical Society and a former chair of the Committee on the Status of Women in Astronomy. She won a teaching award from Rensselaer Polytechnic Institute, a service award from Gallaudet University, and a research award from the University of Memphis. She gives talks and writes articles on topics such as unconscious bias, stereotype threat, and the gender gap. She was honored in 2015 as one of NATURE's top ten people who made a difference in science for her work fighting sexual harassment. 
\end{IEEEbiography}

\begin{IEEEbiography}  [{\includegraphics[width=1in,height=1.25in,clip,keepaspectratio]{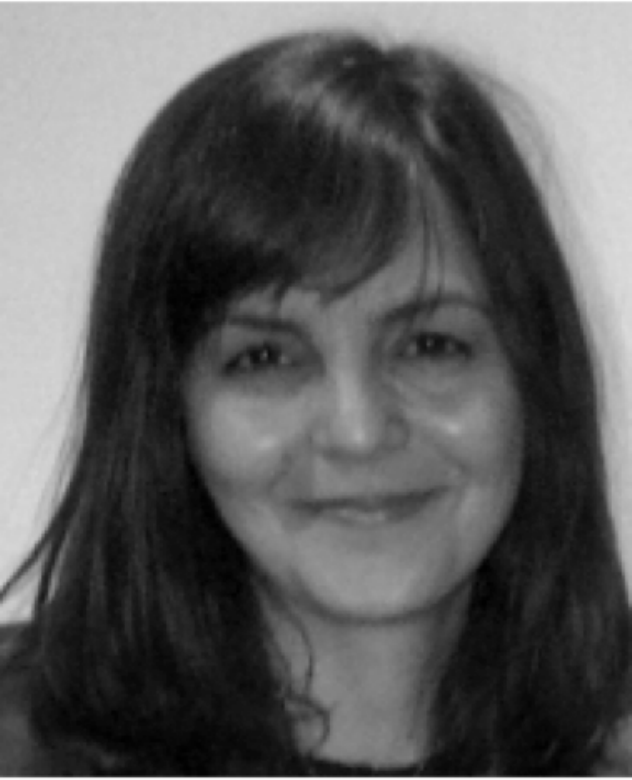}}]{Mahboubeh Asgari-Targhi}
received her BSc degree from University College London, UK. She received her MSc from Imperial College, London, UK and her PhD from University College London. She is currently an Astrophysicist working at the Harvard-Smithsonian Center for Astrophysics. Her primary research areas include solar physics, plasma physics, plasma astrophysics with particular emphasis on the solar corona and solar wind, and applications of topology and geometry in astrophysics.
\end{IEEEbiography}

%\newpage

% on the last page and whether or not the colu\includegraphics[]

%\vfill


\begin{table*}																		
\centering																		
\renewcommand{\arraystretch}{1.0}																		
\caption{Source data}Ê																
\begin{tabular}{|*{8}{p{1.2cm}|}}																		
\hline																		
RA	&	Declination	&	Longitude	&	Latitude		&	No. of	&	No. of	&	Average.	&	Standard	\\Ê	
 ($^\circ$)	&	($^\circ$)	&	($^\circ$)	&	($^\circ$)   &	profiiles	&	Gaussians	&	$\tilde{\chi}^{2}$	&	deviation	\\	
\hline																		
{(1)}	&	{(2)}	&	{(3)}	&	{(4)}	&	{(5)}	&	{(6)}	&	{(7)}	&	{(8)}		\\Ê	 
357.0	&	10.6	&	98.6	&	-49.3	 	&	14	&	118	&	1.12 	&	0.18	\\
357.6	&	9.6	&	100.5	&	-50.8		&	13	&	108	&	1.17	&	0.23	\\
358.0	&	10.3	&	99.9	&	-49.9		&	22	&	107	&	1.10	&	0.19	\\
358.7	&	9.6	&	100.5	&	-50.8 	&	14	&	121	&	1.23	&	0.37	\\
359.0	&	9.5	&	93.8	&	-49.1		&	15	&	131	&	1.06	&	0.17	\\
360.0	&	8.7	&	102	&	-52.1		&	15	&	121	&	1.10	&	0.15	\\
\hline
0.1	&	12	&	103.7	&	-49.0		&	22	&	211	&	1.05	&	0.15\\
0.6	&	12	&	104.4	&	-49.1		&	14	&	127	&	1.04	&	0.19\\
1.1	&	12	&	105.2	&	-49.2		&	13	&	125	&	1.10	&	0.17\\
1.6	&	12	&	105.9	&	-49.4		&	13	&	126	&	1.12	&	0.18\\
2.1	&	12	&	106.6	&	-49.5		&	13	&	118	&	1.12	&	0.16\\
2.6	&	12	&	107.3	&	-49.6		&	10	&	83	&	1.07	&	0.08	\\													
\hline				
\end{tabular}Ê																		
\end{table*}

\begin{table*}																		
\centering																		
\renewcommand{\arraystretch}{1.0}																		
\caption{Derived parameters }Ê																
\begin{tabular}{|*{6}{p{1.4cm}|}}																		
\hline																		
RA	&	 Total N$_{H}$	  &	Column density 	&	Average line width	&	St.dev. line width	&	Apparent Helium	\\Ê	
 ($^\circ$)	&	sum of profiles (10$^{18}$ cm$^{-2}$	) &	(34 km s$^{-1}$ comps.)	&	(34 km s$^{-1}$ comps.) &  (34 km s$^{-1}$  comps.)  & abundance		\\	
\hline																																				
{(1)}	&	{(2)}	&	{(3)}	&	{(4)}	&	{(5)}	&	{(6)}		\\
357.0	&	8099	&	1110	&	34.1	&	1.5	&	0.137	\\						
357.6	&	7902	&	794	&	34.6	&	0.6	&	0.100	\\						
358.0	&	15081	&	1289	&	33.8	&	0.9	&	0.085	\\						
358.7	&	10385	&	951	&	34.3	&	1.6	&	0.092	\\						
359.0	&	10388	&	725	&	34.5	&	1.2	&	0.070	\\						
360.0	&	9691	&	1135	&	33.9	&	1.5	&	0.117	\\						
\hline																	
0.1	&	15925	&	1774	&	32.9	&	0.9	&	0.111\\							
0.6	&	8647	&	898	&	34.5	&	1.5	&	0.104	\\						
1.1	&	8825	&	642	&	34.3	&	1.5	&	0.073\\							
1.6	&	8880	&	793	&	34.7	&	1.1	&	0.089\\							
2.1	&	7619	&	686	&	34.2	&	1.4	&	0.090\\							
2.6	&	6563	&	471	&	34.6	&	1.8	&	0.072\\
\hline				
\end{tabular}Ê																		
\end{table*}																		
\clearpage

\end{document}